\documentclass[multphys,vecphys]{svmult}

\usepackage{makeidx}         
\usepackage{graphicx,psfrag,amsmath,color,amssymb}
\usepackage{multicol}        
\usepackage{cite}            
\usepackage[bottom]{footmisc}

\makeindex             

\begin{document}

\title{Unconventional Density Waves in Organic Conductors and in Superconductors}
\titlerunning{Unconventional Density Waves...}
\author{Kazumi Maki \and Bal\'azs D\'ora \and Attila Virosztek}
\institute{
Department of Physics and Astronomy, University of Southern
California, Los Angeles CA 90089-0484, USA and 
Max Planck Institute for the Physics of Complex Systems, 
N\"othnitzer Str. 38, D-01187, Dresden, Germany
 \texttt{kmaki@usc.edu}\and Department of Physics, Budapest University of Technology and
Economics, H-1521 Budapest, Hungary
\texttt{dora@kapica.phy.bme.hu}
\and 
Department of Physics, Budapest University of Technology and 
Economics, H-1521 Budapest, Hungary and 
Research Institute for Solid State Physics and Optics, P.O.Box
49, H-1525 Budapest, Hungary
\texttt{viro@szfki.hu}}

\maketitle

Unconventional density waves (UDW) are one of the ground states in metallic crystalline solids
and have been speculated already in 1968. 
However, more focused studies on UDW started only recently, perhaps after the identification
of the low temperature phase in $\alpha$-(BEDT-TTF)$_2$KHg(SCN)$_4$ as unconventional
charge density wave (UCDW) in 2002.
More recently, the metallic phase of Bechgaard salts (TMTSF)$_2$X with X=PF$_6$ and ReO$_4$
under both pressure and magnetic field appears to be unconventional spin density wave (USDW).
The pseudogap regime of high $T_c$ superconductors LSCO, YBCO, Bi2212 and
the one in CeCoIn$_5$ 
belong to d-wave density waves (d-DW).

In these identifications, the angular dependent magnetoresistance and the
giant Nernst 
effect have played the crucial role. These are the simplest manifestations
of the Landau quantization of quasiparticle energy in UDW in the presence of magnetic
field (the Nersesyan effect). Also we speculate
that UDW will be most likely found in $\alpha$-(BEDT-TTF)$_2$I$_3$, 
$\alpha$-(BEDT-TTF)$_2$I$_2$Br, $\kappa$-(BEDT-TTF)$_2$Cu(NCS)$_2$, 
$\kappa$-(BEDT-TTF)$_2$Cu(CN)$_2$Br, $\lambda$-(BEDT)$_2$GaCl$_4$ and in many other
organic compounds.

\section{Introduction}
\label{sec:1}

Until very recently, the electronic ground state in crystalline solids were considered to belong to 
one of four canonical ground states in quasi-one-dimensional systems: s-wave
superconductors, p-wave 
superconductors, (conventional) charge density waves and spin density waves\cite{solyom,jerome,gruner,ishiguro}.
Indeed many condensates discovered since 1972 have appeared to be accommodated comfortably
in this scheme. For example the two CDW in NbSe$_3$ and SDW in (TMTSF)$_2$PF$_6$ are well known 
examples\cite{gruner,ishiguro}.

In all these systems, the elementary excitations are of Fermi liquid nature
\'a la Landau\cite{landau1,landau2,landau3} but have the 
energy gap $\Delta$ and the quasiparticle (condensate) density 
decreases exponentially like $\exp(-\Delta/T)$ for $T\le T_c/2$, where
$T_c$ is the transition temperature. Therefore only a small portion of the
quasiparticles will be left below $T_c/2$.

The thermodynamics of these systems are practically the same as those in
s-wave superconductors as described by the theory of Bardeen, Cooper and
Schrieffer (the BCS theory\cite{BCS}). 

However, since the discovery of heavy fermion superconductors
CeCu$_2$Si$_2$\cite{steglich}, organic superconductors in Bechgaard salts
(TMTSF)$_2$PF$_6$\cite{tmtsfexp}, high $T_c$ cuprate superconductors\cite{bednorz}, Sr$_2$RuO$_4$\cite{mackenzie} etc.,
this simple scheme has to be necessarily modified. First of all, most of
these superconductors are unconventional and nodal\cite{sigrist,revmod,damas,makipssb,vanyolos}. Parallel to this
development, there is a surge of studies on CDW whose quasiparticle energy
gap $\Delta(\bf k)$ has line or point nodes at the Fermi surface\cite{capel,benfatto,nagycikk,nayak}. We may
characterize this as the paradigm shift from  one dimensional to higher
dimensional physics\cite{mplb}. However, such paradigm shift been already
anticipated. The quasiparticle excitations are of Fermi liquid type and not
of Tomonaga-Luttinger   type\cite{ishiguro,mihaly2,zhelez}. Note that we are not interested in the high
temperature behaviour $T>200$~K of Bechgaard salts currently discussed in
literature\cite{giamarchi}. 
In higher dimensions, the imperfect nesting is inevitable. Further in a magnetic field,
the quasiparticle spectrum is quantized \'a la Landau
in both SDW\cite{bojana} and CDW\cite{mcdonald}.
This gives rise to
field induced spin density wave (FISDW) with integral quantum Hall effect\cite{ishiguro,vuletic,yak}.

Unconventional density waves were first speculated by Halperin and Rice\cite{HR} as
a possible ground state in the excitonic insulator. Unlike conventional DW,
there is no X-ray or spin signal due to the zero average of the gap over
the Fermi surface ($\langle\Delta({\bf k})\rangle=0$). Therefore UDW is
often called as the state with hidden order parameter. However, certain
confusion has spread around in recent literature with the use of "hidden order".
Therefore it is much wiser to specify what one means. In this sense we
prefer to use the notation UDW. Also unlike many other people\cite{capel,benfatto,nayak,affleck,thalmeier}, we do not
consider the discretized lattice and UDW with minuscule loop current or
$Z_2$ symmetry breaking common to the descendent of the flux phase\cite{affleck,thalmeier}.

It is easy to see that such two dimensional solution is unstable in three
dimensional environment.
In contrast, our UDW has the $U(1)$ gauge symmetry as in conventional
DW. Therefore our UDW can slide, and exhibits the nonohmic conductance\cite{ltp,rapid} as
in conventional DW and generates the phase vortices.

As an other difference compared to conventional DW, the nodal excitations
persist down to $T=0$~K, giving rise to an electronic specific heat $\sim
T^2$.
Indeed, the thermodynamics of UDW is identical to the one in d-wave
superconductors\cite{nagycikk,d-wave}.
The transition from the normal metallic state to UDW is metal-to-metal. As an
example, the $T^2$ dependence of the electric resistance in the normal
state typical to Fermi liquids changes to $T$ linear dependence for $T\le
T_c/2$, which is to be contrasted to the exponentially activated behaviour
in the conventional counterpart.

In light of the discussion above, many of the so-called non-Fermi liquids
can be UDW, and in fact Fermi liquids as defined by Landau, as we shall
see. The change in the exponents distinguishing between various phases
stems from the change in the excitation spectrum, and not from the
different nature of the elementary excitations.
We have shown already that the pseudogap phases in both high $T_c$ cuprates
like LSCO, YBCO and Bi2212\cite{capnernst,dorauj} and CeCoIn$_5$\cite{cecoin,hu} are d-wave density waves based on
the giant Nernst effect and the angle dependent magnetoresistance observed
in these systems\cite{wangxu,capan,wangong,sandu,bel}.
Through the angle dependent photoemission spectra, the d$_{x^2-y^2}$
symmetry of the energy gap in the pseudogap region of high $T_c$ cuprates
has been established\cite{ding}. Also the presence of Fermi arc around the $(\pi,\pi)$
direction\cite{campusano} is consistent with d-DW\cite{gossamer}. Furthermore, the mysterious relation
$\Delta(0)\simeq 2.14 T^*$ found in the pseudogap region of LSCO, YBCO and
Bi2212\cite{nakano,oda,kugler,sutherland} can readily be interpreted in terms of UDW. Here $\Delta(0)$ is the
maximal energy gap determined by STM and $T^*$ is the pseudogap crossover
temperature, which may be identified as $T_c$ in d-DW. The 2.14 gap 
maximum-transition temperature ratio is
the weak-coupling theory value for d-DW and for d-wave superconductors\cite{nagycikk,d-wave}.

\section{Mean-field theory}

In the followings we limit ourselves to quasi-one and quasi-two-dimensional
systems, since for UDW\\
a. the Fermi surface nesting plays an important role in the realization of
the phase,\\
b. we do not have yet any well established examples in three
dimensional systems.\\
We consider the effective (low energy) Hamiltonian given by
\begin{equation}
H=\sum_{\bf k,\sigma}\xi({\bf k})c_{\bf k,\sigma}^{+}c_{\bf
 k,\sigma}+ \frac{1}{2} \sum_{\begin{array}{c}
          {\bf k,k^\prime,q} \\
          \sigma,\sigma^\prime
         \end{array}} {V}({\bf k,k^\prime,q})c_{\bf k+q,\sigma}^{+}
         c_{\bf k,\sigma}c_{\bf k^\prime-q,\sigma^\prime}^{+}c_{\bf
         k^\prime,\sigma^\prime},\label{ham}
\end{equation}
where $c_{\bf k,\sigma}^+$  and
 $c_{\bf k,\sigma}$ are the creation and annihilation operators of electrons with momentum $\bf k$ and spin $\sigma$, $\xi({\bf k})$ is 
the kinetic energy of electrons measured from the Fermi energy in the
 normal state and $V(\bf k,k^\prime,q)$ is the interaction between 
two particles. In the following we represent $V(\bf k,k^\prime,q)$ by a separable potential
\begin{equation}
V({\bf k,k^\prime,q})=\langle|f({\bf k})|^2\rangle^{-1}f({\bf k})f({\bf k^\prime})\delta(\bf q-Q),
\end{equation}
where $\bf Q$ is the nesting vector. 
This kind of interaction is readily obtained from the extended Hubbard
model with at most nearest-neighbour interaction (on site+nearest neighbour
Coulomb interaction, exchange, pair hopping and assisted hopping
terms \cite{nagycikk}). For quasi-one-dimensional systems with most conducting or chain
direction parallel to the $x$ axis, we consider $f({\bf k})=\cos(bk_y)$ or
$\sin(bk_y)$.
For quasi-two-dimensional systems, we consider only d-wave density waves
with d$_{x^2-y^2}$ and d$_{xy}$ symmetry. At present, these four UDW
appears to exhaust all known cases. The quasi-one-dimensional description
is applicable to both 
 $\alpha$-(BEDT-TTF)$_2$KHg(SCN)$_4$ and (TMTSF)$_2$PF$_6$\cite{epladmr,admrprl,kang2,tmtsf}, while
quasi-two-dimensional UDW applies to the pseudogap phase in high $T_c$
cuprates and in CeCoIn$_5$\cite{capnernst,dorauj,cecoin,hu}.

\begin{figure}[h!]
\centering
\psfrag{x}[t][b][1][0]{$E/\Delta$}
\psfrag{y}[b][t][1][0]{$N(E)/N_0$}
{\includegraphics[width=6cm,height=6cm]{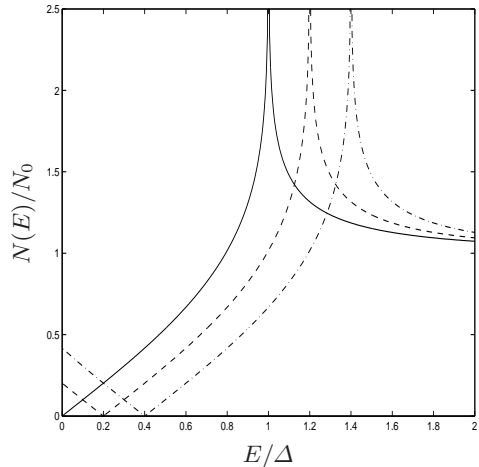}}
\caption{The quasiparticle density of states of UDW is shown for $\mu/\Delta=0$ (solid line), 0.2 (dashed line) and 
0.4 (dashed-dotted line).
\label{dosmu}}
\end{figure}

\begin{figure}[h!]
\centering
\psfrag{x}[t][b][1][0]{$\Gamma/\Gamma_c$}
\psfrag{y}[b][t][1][0]{$\sigma_{yy}(0)/\sigma{yy}^n(0)$}
{\includegraphics[width=6cm,height=6cm]{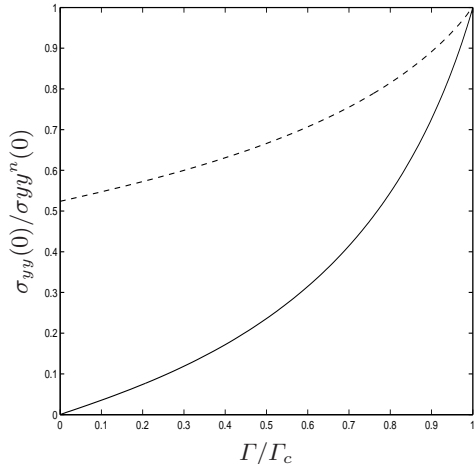}}
\caption{The dc conductivity normalized by its normal state value at $T=0$~K is plotted in the unitary scattering limit 
for $\Delta({\bf k})\sim \sin(bk_y)$ 
(solid line) and $\cos(bk_y)$ (dashed line). $\Gamma_c$ is the critical scattering rate, where the DW phase vanishes.
Similar curves are obtained for the Born limit.
\label{dcunit}}
\end{figure}

Within the mean-field approximation, the Hamiltonian in Eq. \eqref{ham} is
reduced to quadratic form as
\begin{equation}
H=\sum_{\bf k,\sigma}\left(\xi({\bf k})c_{\bf k,\sigma}^{+}c_{\bf k,\sigma}+\Delta({\bf k})c_{\bf k,\sigma}^{+}c_{\bf 
k-Q,\sigma}+\overline{\Delta}({\bf k})c_{\bf k-Q,\sigma}^{+}c_{\bf k,\sigma}\right)-\sum_{\bf k}\frac{|\Delta({\bf 
k})|^2}{V\langle|f({\bf k})|^2\rangle}
\end{equation}
and $\Delta(\bf k)$ obeys the self-consistency equation:
\begin{equation}
\Delta({\bf k})=Vf({\bf k})\sum_{\bf k^\prime,\sigma}f({\bf k^\prime})\langle a_{\bf k^\prime-Q,\sigma}^{+}a_{\bf 
k^\prime,\sigma}\rangle.
\end{equation}
The Hamiltonian is further rewritten as
\begin{equation}
H=\sum_{\bf k,\sigma}\Psi^+_\sigma({\bf k})\left[\tilde\xi({\bf
k})\rho_3+\eta({\bf k})+\Delta({\bf k})\rho_1\right]\Psi_\sigma({\bf k})
-\sum_{\bf k}\frac{|\Delta({\bf k})|^2}{V\langle|f({\bf k})|^2\rangle},
\end{equation}
where $\tilde\xi({\bf k})=(\xi({\bf k})-\xi({\bf k-Q}))/2$ and 
$\eta({\bf k})=(\xi({\bf k})+\xi({\bf k-Q}))/2$.
in the following we shall take the tilde off $\tilde\xi({\bf k})$,
$\eta({\bf k})$ is the imperfect nesting term.
From now on we limit ourselves to UCDW for simplicity. For USDW
we have to involve the Pauli spin matrices as well, though the parallel
treatment is possible. Also $\Psi^+_\sigma({\bf k})$ and $\Psi_\sigma({\bf
k})$ are spinor fields conjugate to each other, $\Psi^+_\sigma({\bf
k})=(c_{\bf k,\sigma}^{+},c_{\bf k-Q,\sigma}^{+})$. Finally the single
particle Green's function or the Nambu-Gor'kov Green's function is given by
\begin{equation}
G^{-1}(\omega,{\bf k})=\omega-\xi({\bf k})\rho_3-\eta({\bf k})-\Delta({\bf k})\rho_1.
\label{green}
\end{equation}
Here $\rho_i$'s are the Pauli matrices operating on the Nambu spinor
space\cite{nambu}. Then the poles of $G(\omega,{\bf k})$ determine the quasiparticle
energies as
\begin{equation}
\omega=\eta({\bf k})\pm\sqrt{\xi({\bf k})^2+\Delta({\bf k})^2}.
\end{equation}
From this, the quasiparticles density of states follows as\cite{nagycikk,d-wave}
\begin{equation}
\frac{N(E)}{N_0}=\textmd{Re}\left\langle\dfrac{|E-\eta({\bf
k})|}{\sqrt{(E-\eta({\bf k}))^2-\Delta^2({\bf k })}}\right\rangle,
\end{equation}
where
$\langle\dots\rangle$ means the average over the Fermi surface, $N_0$ is
the quasiparticle density of states in the normal state at the Fermi
energy.
When $\eta({\bf k})=0$, all UDW so far discussed acquire the same density
of states $N(E)=N_0g(E/\Delta)$ with
\begin{eqnarray}
g(x)=\left\{\begin{array}{l}
\dfrac 2\pi x K(x) \textmd{ for } x<1\medskip\\
\dfrac 2\pi  K(x^{-1}) \textmd{ for } x>1
\end{array}
\right.,
\end{eqnarray}
where $K(x)$ is the complete elliptic integral of the first kind\cite{d-wave}.
For d-wave density waves, in most cases $\eta({\bf k})=\mu$, i.e. the
inclusion of a finite chemical potential as imperfect nesting
suffices. Then we obtain
\begin{equation}
N(E)=N_0g\left(\frac{E-\mu}{\Delta}\right).
\end{equation}
For nonzero chemical potential, $N(0)/N_0=g(\mu/\Delta)\neq 0$. This is
shown in Fig. \ref{dosmu}.
Therefore the chemical potential produces the Fermi arc or
the Fermi pockets detected in ARPES\cite{campusano}.
In quasi-one-dimensional systems, the difference between the different gap
functions $f(\bf k)$ is most readily seen in the optical conductivity
$\sigma_{yy}(\omega)$ (i.e. the one with electric current parallel to the
Fermi surface)\cite{imperfect}. The universal electric conduction (in analogy to the
universal heat conduction in nodal superconductors\cite{vanyolos}) implies that
$\sigma_{yy}(0)\rightarrow 0$ as $\Gamma$ and $T\rightarrow 0$ for
$\Delta({\bf k})\sim \sin(bk_y)$, while $\sigma_{yy}(0)\rightarrow
2e^2N_0v_y^2/\pi\Delta(0)$ for   $\Gamma$ and $T\rightarrow 0$ for 
$\Delta({\bf k})\sim \cos(bk_y)$\cite{doraujimp}. Here $\Gamma$ is the quasiparticle
scattering rate in the normal state. The $\Gamma$ dependence of
$\sigma_{yy}$ is shown in Fig. \ref{dcunit}. Note that the electric conductivity
increases with quasiparticle scattering, which is very counter-intuitive.
But this has a similar origin to the universal heat conduction in nodal
superconductors.
In quasi-two-dimensional systems, the question of d$_{x^2-y^2}$-wave
density wave versus d$_{xy}$-wave density wave is most easily decided by
the angle dependent magnetoresistance, when the magnetic field is rotated
within the conducting plane as we shall discuss in Section \ref{secadmr}\cite{dorauj}.

\section{Landau quantization}
\label{lanquant}

In the presence of a magnetic field, the quasiparticle energy in UDW is
quantized as first discussed by Nersesyan et al. \cite{Ners1,Ners2}. The quasiparticle spectrum
in the presence of magnetic field is obtained from
\begin{equation}
\left(E-\xi({\bf k}+e{\bf A})\rho_3-\eta({\bf k}+e{\bf A})-\Delta({\bf k}+e{\bf A})\rho_1\right)\Psi({\bf r})=0,
\label{dirac}
\end{equation}
where we have introduced the magnetic field through the vector potential
$\bf A$ ($\bf B=\nabla\times A$). 
It is readily recognized that Eq. (\ref{dirac}) has 
the same mathematical structure as the Dirac equation in a magnetic field studied in 1936\cite{heisenberg,weisskopf}.
Since the Landau quantization in quasi-one-dimensional UDW has been
throughoutly discussed in Ref. \cite{mplb} , here we consider the Landau quantization
in d-wave density waves\cite{dorauj}. Without loss of generality, we consider here
d$_{xy}$-wave DW or $\Delta({\bf k})=\Delta\sin(2\phi)$. We assume that
the nodal lines are perpendicular to the conducting plane and run parallel
to the $z$ axis at $(\pm k_F,0,0)$ and $(0,\pm k_F,0)$. Also we take 
$\xi({\bf k})=v(k_\parallel-k_F)+\frac{v'}{c}\cos(ck_z)$, where
$k_\parallel$ is the radial vector within the a-b plane, $v$ and $v'$ are
the respective
 Fermi velocities in the a-b plane and parallel to the c-axis.

Let us assume that a magnetic field lies in the $z-x'$ plane and tilted by
an angle $\Theta$ from the $z$-axis, 
$x'=\hat x\cos{\phi}+\hat y \sin(\phi)$:
\begin{equation}
{\bf B}=B(\cos(\Theta)\hat z +\sin(\Theta)(\cos(\phi)\hat x+\sin(\phi))).
\end{equation}
We shall focus on the quasiparticle spectrum in the vicinity of Dirac cones
at $(\pm k_F,0,\pm \pi/2c)$ and $(0,\pm k_F,\pm \pi/2c)$. Then it is
convenient to choose the vector potential as
\begin{equation}
{\bf A}=-B(\hat z\sin(\Theta)+\cos(\Theta)(\hat x\cos(\phi)+\hat y\sin(\phi)))(y\cos(\phi)-x\sin(\phi)).
\end{equation}
Then in the vicinity of Dirac cones, Eq. \eqref{dirac} is recasted as
\begin{gather}
\left[E-\mu+eB(x\sin(\phi)-y\cos(\phi))(v\cos(\Theta)\cos(\phi)\pm
v'\sin(\Theta))\rho_3-\right.\nonumber\\
\left.
v_2(-\textmd{i}\partial_y)\rho_1\right]\Psi(\bf r),
\end{gather}
where $v_2=2\Delta/k_F$. From this, we obtain
\begin{gather}
\left(E_{1n\pm}-\mu\right)^2=2neBv_2|\cos(\phi)(v\cos(\Theta)\cos(\phi)\pm
v'\sin(\Theta))|\label{specegy}\\
\left(E_{2n\pm}-\mu\right)^2=2neBv_2|\sin(\phi)(v\cos(\Theta)\sin(\phi)\pm
v'\sin(\Theta))|\label{specketto}
\end{gather}
and $n=0$, 1, 2, 3\dots.

Except for the $n=0$ state, which is nondegenerate, all other states are doubly degenerated. Also unlike
in UDW in quasi-one dimensional systems, the Landau spectrum consists of four different branches. Furthermore, the 
chemical potential $\mu(\neq 0)$ is crucial in interpreting the magnetotransport of quasiparticles\cite{dorauj,hu}. 
Finally, Eq. 
\eqref{specegy} gives the particle and hole spectrum
\begin{equation}
E_{1n-}=\mu\pm\sqrt{2neBv_2|\cos(\phi)(v\cos(\Theta)\cos(\phi)-v'\sin(\Theta))|}.
\label{alap}
\end{equation}
A similar formula can be worked out for Eq. \eqref{specketto}. From these 
Landau spectra, the thermodynamics as well as the transport properties are readily obtained. In the following, we shall 
consider the angle dependent magnetoresistance, the non-linear Hall constant and the giant Nernst effect as the three 
hallmarks of UDW.

\section{Angle dependent magnetoresistance (ADMR)}
\label{secadmr}

\subsection{$\alpha$-(BEDT-TTF)$_2$MHg(SCN)$_4$ salts with M=K, Rb and Tl}

The nature of the low temperature phase (LTP) in the quasi-two-dimensional organic conductor 
{$\alpha$-(BEDT-TTF)$_2$MHg(SCN)$_4$ salts with M=K, Rb and Tl} has not been understood until recently\cite{singleton}.
Although the phase transition is clearly seen in magnetotransport measurements, neither charge not magnetic order 
has 
been established\cite{jetp,karts1}. Moreover, the destruction of the LTP in an applied magnetic field suggests a kind of 
CDW rather 
than 
SDW. But the threshold electric field associated with the sliding motion of the density wave turns out to be very 
different from the one in CDW, but somewhat similar to that observed in SDW\cite{tomic}: the threshold field increases 
smoothly with 
temperature and does not diverge at 
$T_c$.
Indeed, the threshold electric field in the LTP of $\alpha$-(BEDT-TTF)$_2$KHg(SCN)$_4$\cite{ltp,rapid,tesla}
is well described in terms of imperfectly nested UCDW. More recently, striking angular dependent magnetoresistance
has been detected in this material\cite{caulfield2,kovalev,fermi,hanasaki}. There have been many unsuccessful attempts 
to interpret this phenomenon in terms 
of the reconstruction of the Fermi surface.

Rather we find that Landau quantization of the quasiparticle spectrum in UDW as described in Sec. \ref{lanquant} plays  
the crucial role\cite{epladmr,admrprl}. Also the giant Nernst effect found in LTP of $\alpha$-(BEDT-TTF)$_2$KHg(SCN)$_4$
can be described in terms of Landau quantized UCDW\cite{choi,nernst}. Therefore these findings lead us to the conclusion 
that the LTP in
$\alpha$-(BEDT-TTF)$_2$MHg(SCN)$_4$
with M=K, Rb and Tl is UCDW.

\begin{figure}[h!]
\centering
\psfrag{a}[t][b][1][0]{$a$}
\psfrag{b}[][][1][0]{$c$}
{\includegraphics[width=4cm,height=4cm]{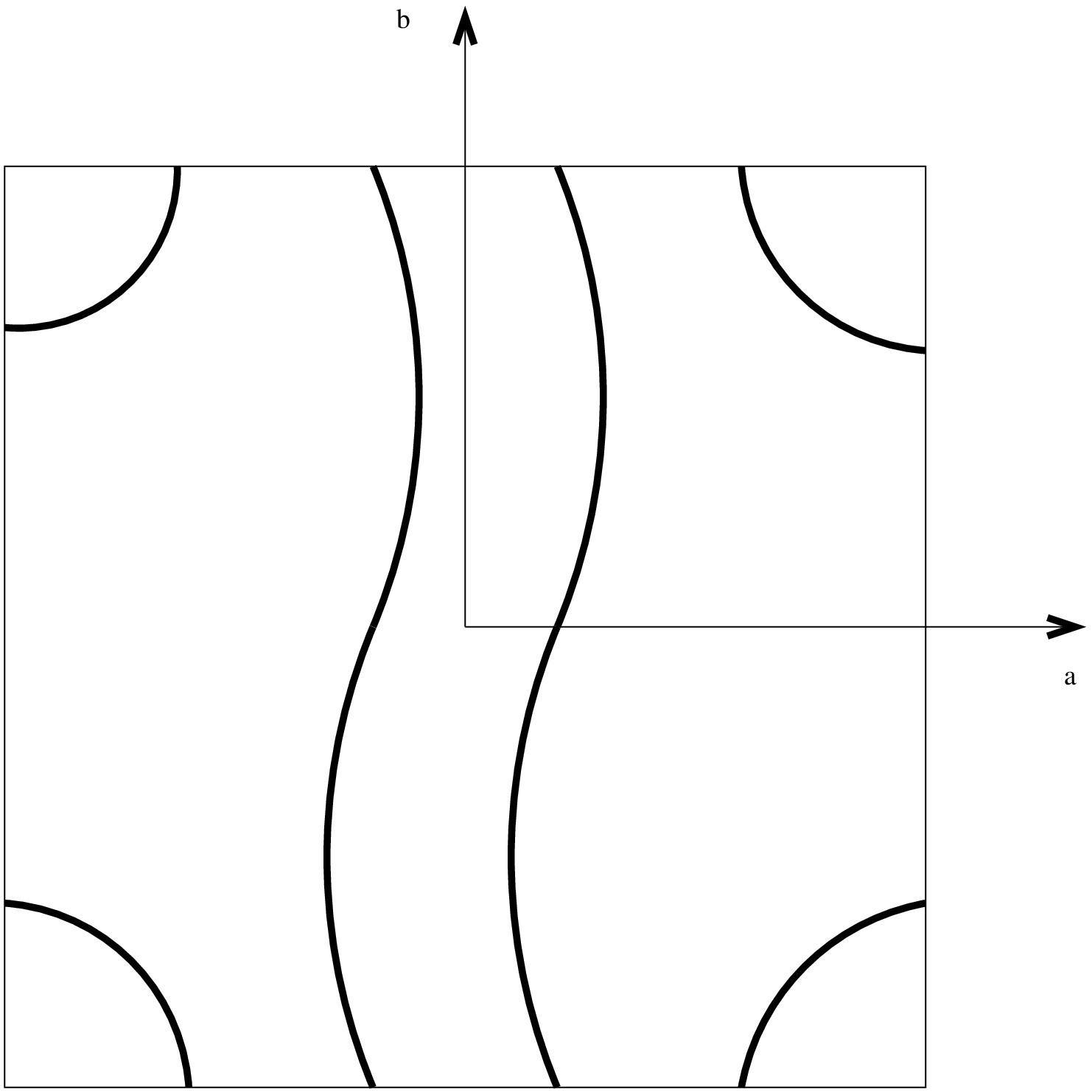}}
\hspace*{10mm}
\psfrag{B}[bl][tr][1][0]{$\bf B$}
\psfrag{c}[b][t][1][0]{$c$}
\psfrag{a}[b][t][1][0]{$a$}
\psfrag{b}[r][l][1][0]{$b$}
\psfrag{pp}[][][1][0]{$\phi$}
\psfrag{p}[][][1][0]{$\theta$}
\includegraphics[width=4cm,height=4cm]{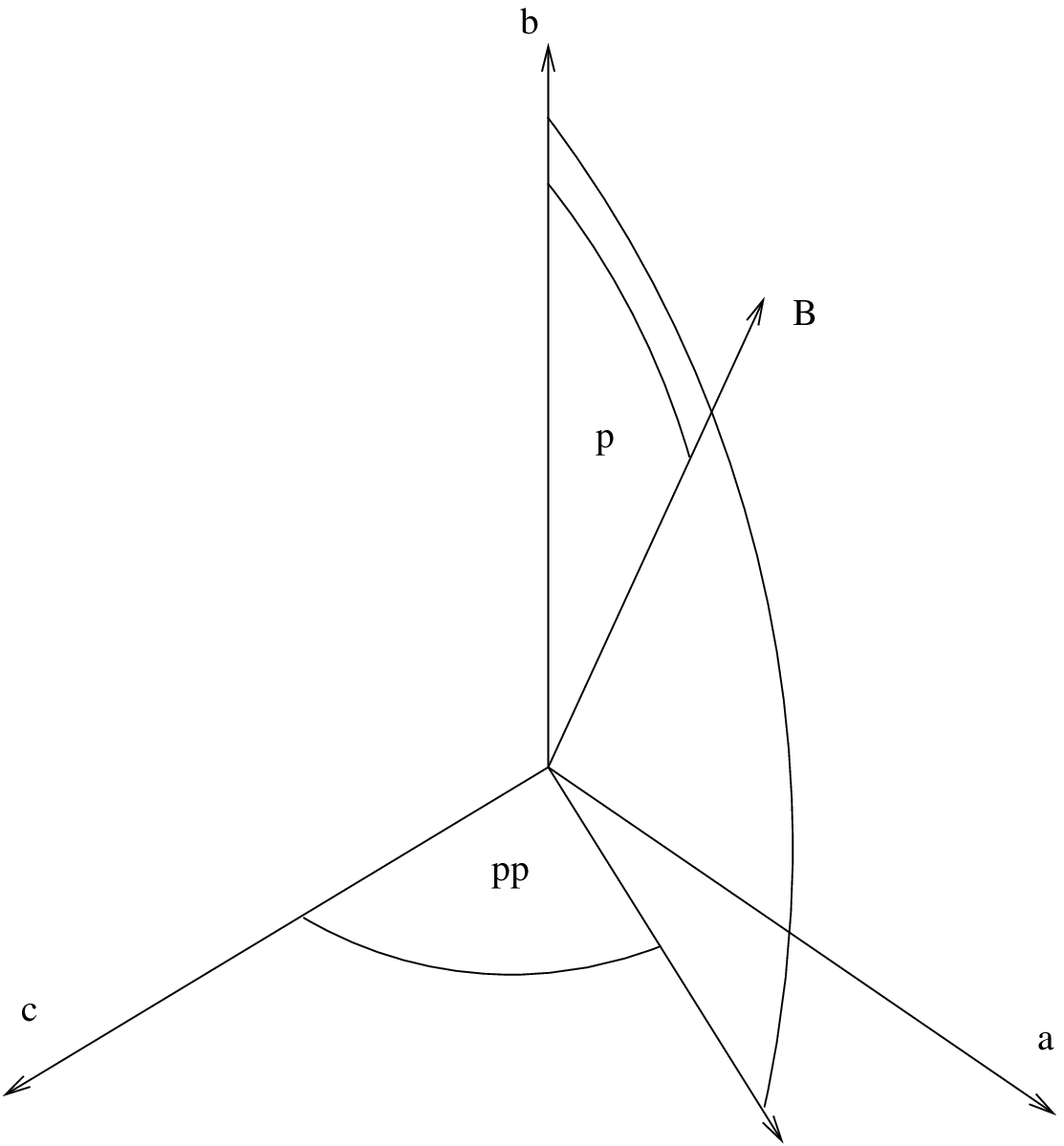}
\caption{The Fermi surface of $\alpha$-(BEDT-TTF)$_2$KHg(SCN)$_4$ is shown schematically in the left panel. 
In the right one the
geometrical
configuration of the magnetic field with respect to
 the conducting plane is plotted.\label{fermisurf}}
\end{figure}

Before proceeding, it is useful to check the Fermi surface of $\alpha$-(BEDT-TTF)$_2$MHg(SCN)$_4$ salts as shown in 
Fig. \ref{fermisurf}\cite{ishiguro,singleton}.

It consists of a quasi-one-dimensional Fermi surface perpendicular to the $a$ axis (the most conducting direction) and 
small two dimensional pockets (quasi-two-dimensional Fermi surface) at the corners of the Brillouin zone. Most of the de 
Haas van Alphen oscillations come from the quasi-two-dimensional Fermi surface, while UCDW appears on the 
quasi-one-dimensional one. The magnetic field configuration with respect to the conducting plane is also shown in Fig. 
\ref{fermisurf}, which was used in ADMR measurement.

\begin{figure}[h!]
\centering
\psfrag{x}[t][b][1][0]{$\theta$ ($^\circ$)}
\psfrag{y}[b][t][1][0]{$R_\parallel(15T,\theta)$ (Ohm)}
\includegraphics[width=6cm,height=6cm]{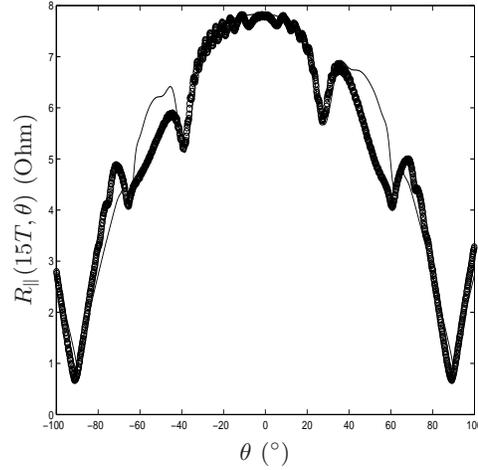}
\caption{The angular dependent magnetoresistance is shown for current parallel
to the a-c plane at $T=1.4$K,
$B=15$T. The open circles belong to the experimental data, the solid
line is our fit based on Eq. \eqref{fit}.}
\label{rpara}
\end{figure}

\begin{figure}[h!]
\centering
\psfrag{x}[t][b][1][0]{$\theta$ ($^\circ$)}
\psfrag{y}[b][t][1][0]{$R_\perp(15T,\theta)$ (Ohm)}    
\includegraphics[width=6cm,height=6cm]{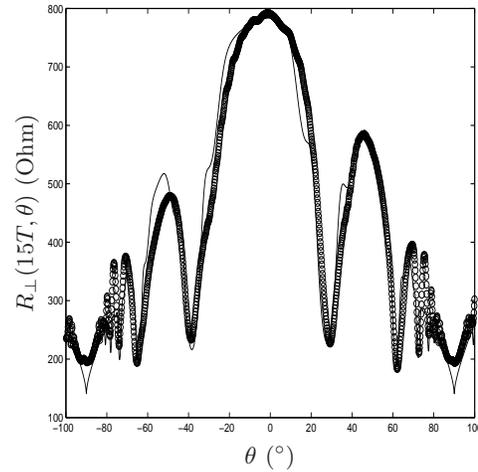}
\caption{The angular dependent magnetoresistance is shown for current perpendicular 
to the a-c plane at $T=1.4$K,
$B=15$T. The open circles belong to the experimental data, the solid
line is our fit from Eq. \eqref{fit}.}
\label{rperp}
\end{figure}

The magnetoresistance is given in terms of the quasiparticle energy $E_n$ as\cite{mplb,dorauj}
\begin{equation}
R^{-1}=\sum_n\sigma_{n}\textmd{sech}^2\left(\frac{\beta E_{n}}{2}\right),
\label{res1}
\end{equation}
where $\sigma_n$'s are the level conductivities weakly depending on temperature and magnetic field, 
$E_n=\sqrt{2nv_a\Delta e|B\cos(\theta)|}$. In the low 
temperature region, where $\beta E_1\gg 1$, Eq. \eqref{res1} is approximated as
\begin{gather}
R^{-1}=2\sigma_0\textmd{sech}^2\left(\frac{\beta E_0^1}{2}\right)+\sigma_1\left[\textmd{sech}^2\left(\frac{\beta 
(E_1+E_1^{(1)})}{2}\right)\right.+\nonumber\\
+\textmd{sech}^2\left(\frac{\beta 
(E_1-E_1^{(1)})}{2}\right)+\textmd{sech}^2\left(\frac{\beta 
(E_1+E_1^{(2)})}{2}\right)+\nonumber\\
\left.+\textmd{sech}^2\left(\frac{\beta 
(E_1-E_1^{(2)})}{2}\right)\right]=\nonumber\\
\frac{4\sigma_0}{1+\cosh(\zeta_0)}+4\sigma_1\left[\frac{1+\cosh(x_1)\cosh(\zeta_0)}{(\cosh(x_1)+\cosh(\zeta_0))^2}+
\frac{1+\cosh(x_1)\cosh(\zeta_1)}{(\cosh(x_1)+\cosh(\zeta_1))^2}\right],
\label{fit}
\end{gather}
where
$\zeta_0=\beta E_0^{(1)}$, $\zeta_1=\beta E_1^{(2)}$ and
$x_1=\beta E_1$, 
\begin{gather}
E_0^{(1)}=E_1^{(1)}=\sum_m\varepsilon_m\exp(-y_m),\\
E_1^2=\sum_m\varepsilon_m(1-2y_m)\exp(-y_m).
\end{gather}
Here $E_0^{(1)}$ and  $E_1^{(1,2)}$ are corrections to Landau level
energies from imperfect nesting (i.e. the warping of the
quasi-one-dimensional Fermi surface):
\begin{equation}
\eta({\bf k})=\sum_n\varepsilon_n\cos(2{\bf
b}^\prime_n\bf k).
\label{generalization}
\end{equation}
From this we find $y_n=v_a {b^\prime}^2e
|B\cos(\theta)|[\tan(\theta)\cos(\phi-\phi_o)-\tan(\theta_n)]^2/\Delta c$,
$\tan(\theta_n)\cos(\phi-\phi_0)=\tan(\theta_0)+nd_0$,
$\tan(\theta_0)\simeq0.5$, 
$d_0\simeq 1.25$, $\phi_0\simeq 27^\circ$\cite{mplb}.
A typical fitting of ADMR of $\alpha$-(BEDT-TTF)$_2$KHg(SCN)$_4$ at
$T=1.4$~K and $B=14$~T is shown in Figs. \ref{rpara} and \ref{rperp}.

\begin{figure}[h!]
\centering
\psfrag{x}[t][b][1][0]{$\theta$ ($^\circ$)}
\psfrag{y}[b][t][1][0]{$R_\perp(15T,\theta)$ (Ohm)} 
\includegraphics[width=5.4cm,height=5.4cm]{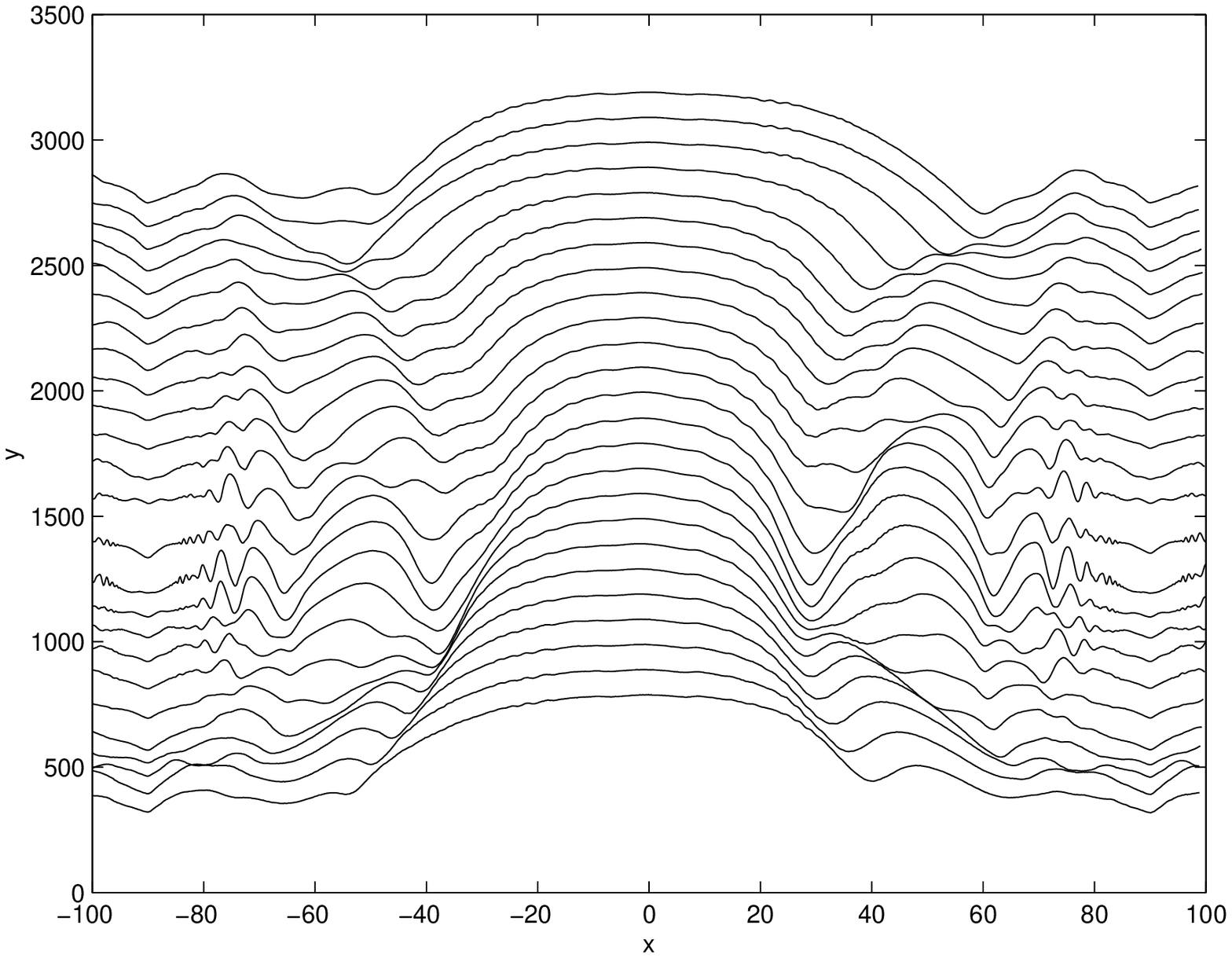}
\hspace*{2mm}
\psfrag{x}[t][b][1][0]{$\theta$ ($^\circ$)}
\psfrag{y}[b][t][1][0]{$R_\perp(15T,\theta)$ (Ohm)} 
\includegraphics[width=5.4cm,height=5.4cm]{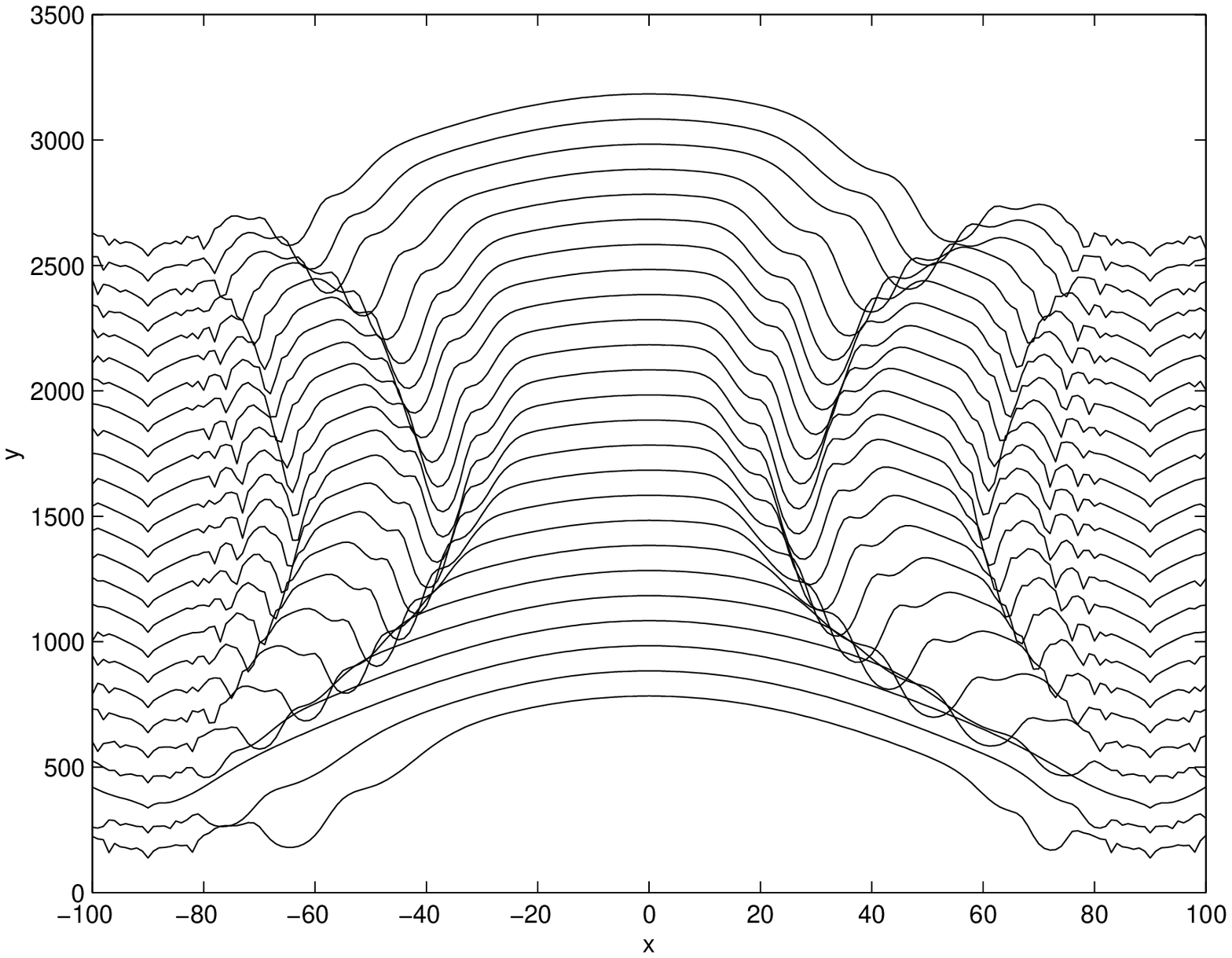}
\caption{ADMR is shown for current perpendicular to the a-c plane at $T=1.4$K and $B=15T$ for $\phi=-77^\circ$, $-70^\circ$,
$-62.5^\circ$, $-55^\circ$, $-47^\circ$, $-39^\circ$, $-30.5^\circ$, $-22^\circ$, $-14^\circ$, $-6^\circ$, $2^\circ$, $10^\circ$,
$23^\circ$, $33^\circ$, $41^\circ$, $48.5^\circ$, $56^\circ$, $61^\circ$, $64^\circ$,
$67^\circ$, $73^\circ$, $80^\circ$, $88.5^\circ$, $92^\circ$ and $96^\circ$ from bottom to top. The left (right) panel shows
experimental
(theoretical) curves, which are shifted from their original position along the vertical axis by $n\times100$Ohm, $n=0$ for $\phi=-77^\circ$, $n=1$ 
for $\phi=-70^\circ$, \dots.}\label{sokphit}
\end{figure}

Broad peaks centered at $\theta=0^\circ$ comes from $E_1$, while the series
of dips stems from the imperfect nesting term. From these fittings, we can
extract
$\Delta$ ($17$K), $v_a\sim 6\times 10^6$cm/s, 
 $b^\prime\sim 30$\AA, $\varepsilon_0\sim 3$K and  $\sigma_2/\sigma_1$ of
the order of $1/10$.
Finally we show in Fig. \ref{sokphit} a series of ADMR data, when the magnetic
field is rotated in the plane as well. Clearly, the theory reproduces the
general features of the ADMR. Of course, we notice that some interesting
details are still missing from our model.
Nevertheless we may conclude that UCDW describes many features of ADMR
observed in $\alpha$-(BEDT-TTF)$_2$KHg(SCN)$_4$\cite{epladmr,admrprl}.

\subsection{Bechgaard salts (TMTSF)$_2$X}

\begin{figure}[h!]
\centering
\psfrag{4}[][][1][0]{4}
\psfrag{8}[][][1][0]{8}
\psfrag{12}[][][1][0]{12}
\psfrag{x}[][][1][0]{$P$~(kbar)}
\psfrag{s}[t][b][1][0]{SDW}
\psfrag{v}[t][b][1][0]{\hspace*{16mm}SDW+USDW}
\psfrag{u}[][][1][0]{USDW}
\psfrag{b}[][][1.5][0]{B=0}
\psfrag{t}[][][1][0]{\hspace*{16mm}triplet SC} 
\psfrag{y}[][][1][90]{$T$}~(K)
\includegraphics[width=8cm,height=5cm]{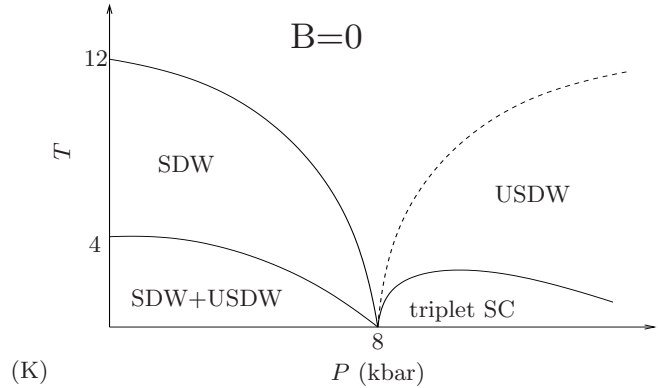}
\caption{The schematic $P-T$ phase diagram of Bechgaard salts.}
\label{tmtphase}  
\end{figure}

The first organic superconductor (TMTSF)$_2$PF$_6$, discovered in 1980\cite{tmtsfexp}, is,
perhaps, one of the most fascinating electron system so far studied.
The Bechgaard salts are well known for the variety of their ground
states. They exhibit spin density wave at ambient to moderate pressure,
field induced spin density wave and triplet superconductivity at high
pressure ($p>8$~kbar)\cite{jerome,ishiguro,triplet} as shown in Fig. \ref{tmtphase}. 
We have discovered recently the coexistence of SDW and USDW at $T=T^*\sim
4$~K in (TMTSF)$_2$PF$_6$ during the analysis of ADMR across $T^*$\cite{basletic}.
Also in the large area of the $P-B$ phase diagram, where both SDW and
superconductivity are suppressed by pressure and magnetic field (the
so-called metallic state), surprising ADMR has been observed about a decade
ago\cite{chaikin}. These are of three kind: first, when the magnetic field is rotated
within the $c^*-b^\prime$ plane perpendicular to the $a$ axis, $R_{xx}$ and
$R_{zz}$ exhibits bread peaks with a number of dips\cite{osada1,naughton,kang1,lee}. These dips were
interpreted in terms of Lebed resonance\cite{lebed1,lebed2}. More recently, very similar ADMR
has been seen in the ReO$_4$ and ClO$_4$ compounds  as well, though the one
in the latter is more complex\cite{kang2}.
Second, when the magnetic field is rotated within the $a-c^*$ plane,
$R_{xx}$ show very different ADMR, sometimes called "Danner resonance"\cite{danner}. Finally the third angular dependence appears as a kink
when the magnetic field is rotated within the $a-b^\prime$ plane.

\begin{figure}[h!]
\centering
\psfrag{x}[t][b][1][0]{$\phi$ ($^\circ$)}
\psfrag{y}[b][t][1][0]{$R_{xx}(\Omega)$}
\includegraphics[width=6cm,height=6cm]{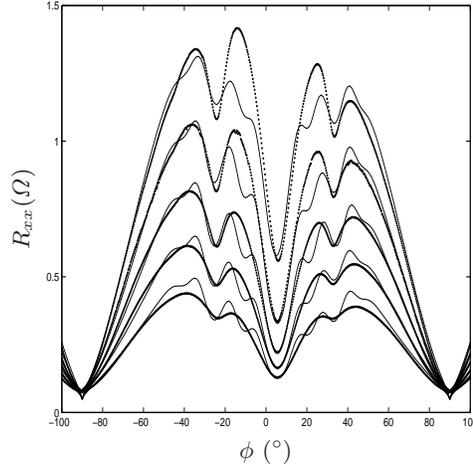}
\caption{The angular dependent magnetoresistance ($R_{xx}$) of (TMTSF)$_2$PF$_6$ at $T=1.55$~K
is
shown for magnetic fields from
8-4~T from top to bottom. The dots denote the experimental data from
\cite{kang2}, the solid line is our fit
based
on equation \eqref{fit}.}
\label{rxx1}  
\end{figure}

\begin{figure}[h!]
\centering
\psfrag{x}[t][b][1][0]{$\phi$ ($^\circ$)}
\psfrag{y}[b][t][1][0]{$R_{xx}(\Omega)$}
\includegraphics[width=6cm,height=6cm]{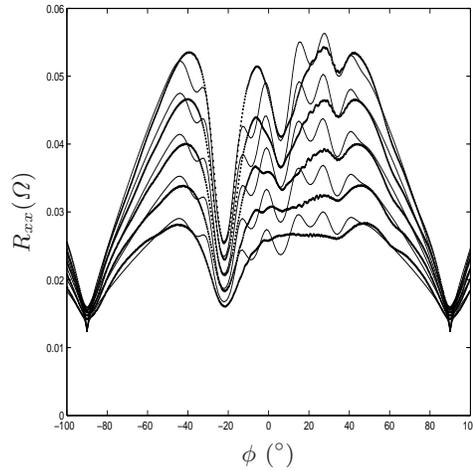}
\caption{The angular dependent magnetoresistance ($R_{xx}$) of (TMTSF)$_2$ReO$_4$ at $T=1.55$~K is
shown for magnetic fields from
8-4~T from top to bottom. The dots denote the experimental data from
\cite{kang2}, the solid line is our fit based 
on equation \eqref{fit}.}
\label{rxx2}
\end{figure}

Here we shall concentrate on the case when the magnetic field is rotated
within the $c^*-b^\prime$ plane. We propose that the metallic phase should
be USDW\cite{tmtsf}. When $\bf B$ in the $c^*-b^\prime$ plane is tilted by $\phi$ from
the $c^*$ axis, the obtained energy spectrum turns out to be very similar
to the one discussed in Eq. \eqref{alap}. In particular, we get
$E_1=\sqrt{2nv_a\Delta e|B\cos(\phi)|}$, 
$y_m=v_a {d^\prime}^2e |B\cos(\phi)|(\tan(\phi)-\tan(\phi_m))^2/\Delta
b$ and 
$\tan(\phi_m)=\frac{pb\sin(\gamma)}{qc\sin(\beta)\sin(\alpha^*)}-\cot(\alpha^*)$,
where $b$, $c$, $\beta$ and $\gamma$ are lattice parameters in real space,
$\alpha^*$ is a lattice parameter in reciprocal space, $p$ and $q$ are small 
integers\cite{tmtsf}. Then the experimental data\cite{kang2} of $R_{xx}$ on (TMTSF)$_2$PF$_6$ and on 
(TMTSF)$_2$ReO$_4$ at
$T=1.55$~K are shown in Figs. \ref{rxx1} and \ref{rxx2}, respectively.
Except for the fact that the theoretical curves exhibit more structures
than the data, we think the fitting is excellent. From these, we extract
$\Delta=20$~K and $45$~K and $v_a=10^7$~cm/s and $3\times 10^7$~cm/s for
the PF$_6$ and ReO$_4$ compound, respectively. 
The Fermi velocities deduced here are also very consistent with known
values.
Also from the above $\Delta$'s, we expect that USDW persists to $T=9$~K and
20~K for the respective compounds. Also the giant Nernst effect in these
systems can be crucial to strengthen the case of USDW in Bechgaard salts\cite{wulee}.

\subsection{$\kappa$-(ET)$_2$ salts, CeCoIn$_5$ and YPrCO}

There are three $\kappa$-(ET)$_2$ salts with very similar properties:
$\kappa$-(ET)$_2$Cu(NCS)$_2$, 
$\kappa$-(ET)$_2$Cu[NCN]$_2$Br and $\kappa$-(ET)$_2$Cu[NCN]$_2$Cl with
relatively high superconducting transition temperature $T_c\gtrsim 10$~K.
So far no evidence for unconventional density wave in these systems has
been reported, although we have many reasons to believe that the pseudogap
phase in high quality crystals should be d-wave density wave. First of all,
There are many parallels between these organic superconductors and high
$T_c$ cuprate superconductors and the heavy fermion compound CeCoIn$_5$:

quasi-two-dimensionality (or layered structure),

the proximity of antiferromagnetic order,

d-wave superconductivity.\\
Of course d-wave superconductivity is established only for
$\kappa$-(ET)$_2$Cu(NCS)$_2$\cite{izawa}, but this suggests strongly that the other two $\kappa$-(ET)$_2$ salt
superconductors should be d-wave as well.
Second, the giant Nernst signal and ADMR observed in the pseudogap phase of high $T_c$ 
cuprates\cite{wangxu,capan,wangong,sandu} and
in CeCoIn$_5$\cite{hu,bel} are successfully interpreted in terms of d-wave density wave\cite{dorauj}. These suggest 
strongly the 
presence of dDW 
in $\kappa$-(ET)$_2$ salts.

\begin{figure}[h!]
\centering
\psfrag{x}[t][b][1][0]{$\theta$}
\psfrag{y}[b][t][1][0]{$\sigma$(1/$\mu\Omega$cm)}
\includegraphics[width=7cm,height=6cm]{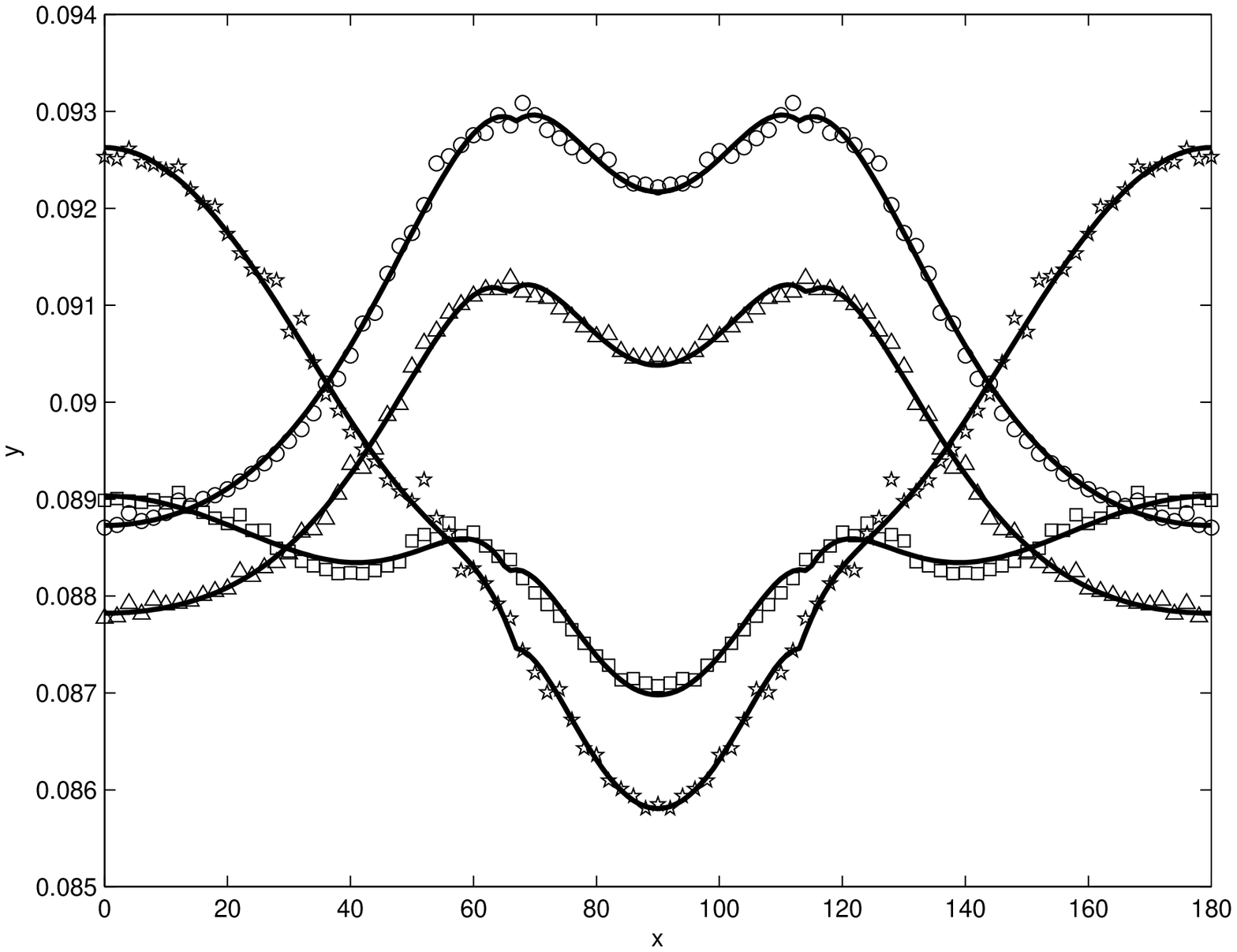}
\caption{The angle dependent conductivity of CeCoIn$_5$ is shown for $T=6$~K and for $B=4$~T (circle), 5~T (triangle),
8~T (square), 10~T (star).
\label{cecoin5}}
\end{figure}

\begin{figure}[h!]
\centering
\psfrag{x1}[t][b][1][0]{$\theta^\circ$}
\psfrag{y1}[][t][1][0]{$\Delta\rho_{ab}/\rho_{ab}$}
\psfrag{x2}[t][b][1][0]{$\theta^\circ$}
\psfrag{y2}[][t][1][0]{$\Delta\rho_{ab}/\rho_{ab}$}
\psfrag{x3}[t][b][1][0]{$\theta^\circ$}
\psfrag{y3}[][t][1][0]{$\Delta\rho_{ab}/\rho_{ab}$}
\psfrag{x4}[t][b][1][0]{$\theta^\circ$}
\psfrag{y4}[][t][1][0]{$\Delta\rho_{ab}/\rho_{ab}$}
\includegraphics[width=10cm,height=7cm]{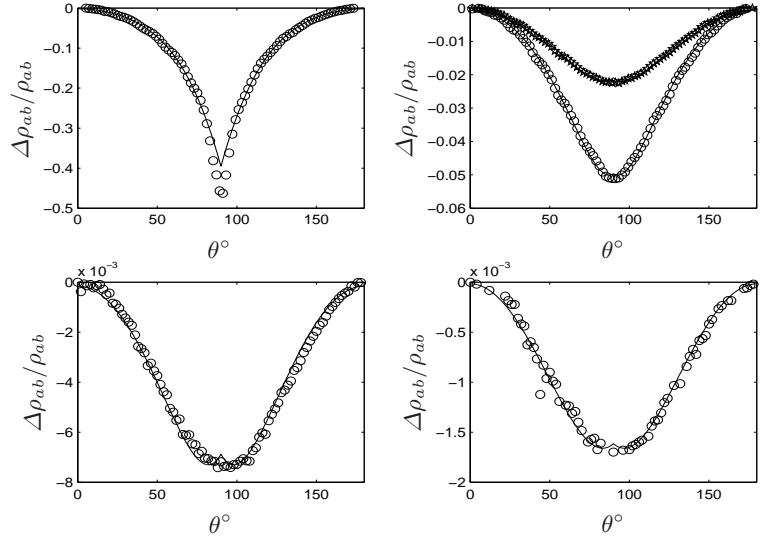}
\caption{The relative change of the in-plane magnetoresistance of
Y$_{0.68}$Pr$_{0.32}$Ba$_2$Cu$_3$O$_7$\cite{sandu} is plotted as a function of angle $\theta$ at B=14~T
for T=52~K (top left), 60~K and 65~K (top right), 75~K (bottom left) and 105~K (bottom right). The solid
line is fit based on Eq. \ref{fit}.}
\label{abra1}
\end{figure}

Here we are going to show the fitting of ADMR in the pseudogap phase of CeCoIn$_5$\cite{hu} and in high $T_c$ cuprate 
Y$_{0.68}$Pr$_{0.32}$Ba$_2$Cu$_3$O$_7$\cite{dorauj,sandu} in Figs. \ref{cecoin5} and \ref{abra1}, respectively.

From these fittings, we can extract $\Delta=45$~K and 360~K, $\mu$(the chemical potential)=8.4~K, 40-60~K, $v$(the 
planar Fermi velocity)=3.3$\times 10^6$~cm/s and 2.3$\times 10^7$~cm/s for CeCoIn$_5$ and YPrCO, respectively. These 
values are very consistent with known parameters of these systems.

\section{Giant Nernst effect}

\begin{figure}[h!]
\centering
\psfrag{x}[t][b][1.2][0]{$B$(T)}
\psfrag{y}[b][t][1.2][0]{$S_{xy}$($\mu$V/K)}
\psfrag{x6}[l][][1][0]{$T=1.4$~K}
\psfrag{x7}[l][][1][0]{$T=4.8$~K}
\includegraphics[width=7cm,height=7cm]{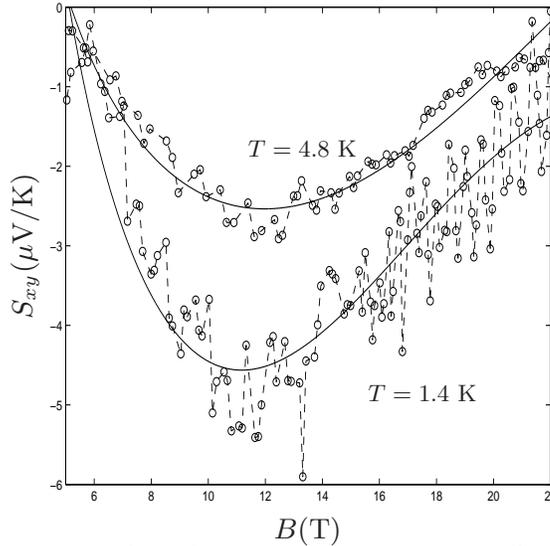}
\caption{The Nernst signal for heat current along the $a$ direction is shown
for $T=1.4$~K and  $T=4.8$~K (from bottom to top), the dashed lines with circles denote the
experimental data from Ref. \cite{choi}, the solid line is our fit based on Eq.
(\ref{nern}).}\label{choi5a}
\end{figure}

Since 2001 the giant Nernst effect has been established as the hallmark of the pseudogap phase in high $T_c$ 
cuprates\cite{wangxu,capan,wangong}. 
As shown in Ref. \cite{mplb,capnernst}, the giant negative Nernst effect follows directly from the Landau quantization 
of the quasiparticle 
energy spectrum in UDW. When an electric field is applied in the conducting plane in addition to the 
perpendicular magnetic field, all quasiparticles drift with a drift 
velocity ${\bf v}_D={\bf E\times B}/B^2$. This gives rise to a heat current ${\bf J}_h=TS{\bf v}_D$, where $S$ is the 
quasiparticle entropy given by
\begin{equation}
S=eB\sum_n\left\{\ln\left(1+e^{-\beta E_n}\right)+\beta E_n\left(1+e^{\beta E_n}\right)^{-1}\right\},
\end{equation}
where the sum has to be carried out over all the Landau levels. For d-wave density wave as in high $T_c$ cuprates and 
for $\bf B\parallel c$ and $\beta E_1\gg 1$, the entropy is well approximated as
\begin{gather}
S=8eB\left[\ln\left(1+e^{-\zeta_0}\right)+\zeta_0\left(1+e^{\zeta_0}\right)^{-1}+\ln\left(2\left(\cosh(x_1)
+\cosh(\zeta_0)\right)\right)-\nonumber\right.\\
\left.-\frac{\zeta_0\sinh(\zeta_0)+x_1\sinh(x_1)}{\cosh(x_1)+\cosh(\zeta_0)}\right],
\end{gather}
where $\zeta_0=\beta\mu$, $x_n=\beta\sqrt{2neBv_2v}$ with $v_2/v=\Delta/E_F$. Then the Nernst coefficient is given by
\begin{gather}
S_{xy}=-\frac{S}{B\sigma}=\frac 1\sigma\left\{\frac{L_{2D}}{1+(B/B_0)^2}\right.-\nonumber\\
\left.-8e\left(\frac 
12\left[\ln(2(1+\cosh(\zeta_0))-\frac{\zeta_0\sinh(\zeta_0)}{1+\cosh(\zeta_0)}\right]\right.\right.+\nonumber\\
\left.\left.
+\ln\left(2\left(\cosh(x_1)
+\cosh(\zeta_0)\right)\right)-\frac{\zeta_0\sinh(\zeta_0)+x_1\sinh(x_1)}{\cosh(x_1)+\cosh(\zeta_0)}\right)\right\}.
\label{nern}
\end{gather}
Here $\sigma$ has been already defined in Eq. \eqref{res1}, and $L_{2D}$  and $B_0$ comes from quasiparticles not in 
d-DW state.
Here we present our fitting to the data taken on $\alpha$-(BEDT-TTF)$_2$KHg(SCN)$_4$\cite{choi} in Fig. \ref{choi5a}.

 We have also similar 
fittings for 
high $T_c$ cuprates\cite{capnernst} and CeCoIn$_5$\cite{cecoin}.
The third hallmark of UDW is the nonlinear Hall coefficient. In UDW in an applied magnetic field, the number of 
quasiparticles decreases exponentially with decreasing temperature. Therefore the Hall conductivity is given 
by\cite{gossamer}
\begin{equation}
\sigma_{xy}=-\frac{2e^2\cos^2(\Theta)}{\pi}n(B,T)
\label{hall}
\end{equation}
with
\begin{eqnarray}
n(B,T)=\tanh(\zeta_0/2)+\nonumber\\
+\sinh(\zeta_0)\left[\frac{1}{\cosh(x_1)+\cosh(\zeta_0)}+
\frac{1}{\cosh(x_2)+\cosh(\zeta_0)}\dots\right],
\end{eqnarray}
where $\Theta$ is the angle the magnetic field makes from the $c$ axis.
We have no possibility to compare the theoretical expression (Eq. \eqref{hall}) to available experimental data.
But clearly the Hall coefficient will indicate the reduction of the number of quasiparticles due to the opening of the 
energy gap.

\section{Concluding remarks}

In the last few years we have experienced a major paradigm shift from conventional condensates to unconventional
condensates. These are still the mean-field ground states and described in terms of the generalized BCS theory.
For superconductivity, at least d-wave superconductivity in $\kappa$-(ET)$_2$ salts and triplet superconductivity 
in (TMTSF)$_2$PF$_6$ (most 
likely chiral  f-wave) has been established\cite{vanyolos,dominguez}. 
As to density waves, more and more previously unidentified condensates appear to belong to UDW. But this game has just 
started. Also there appears to be still unexplored areas where two of the mean-field order parameters can coexist. Of 
course, the classic example is NbSe$_3$ where the CDW's: CDW$_1$ and CDW$_2$ coexists. A more exotic case will be in 
Bechgaard salts (TMTSF)$_2$PF$_6$ where SDW and USDW appears to coexist below $T^*\sim T_c/3$\cite{basletic}.
However, Gossamer superconductivity appears to be more widespread, where a nodal superconductor coexists with a nodal 
density wave\cite{gossamer,laughlin}. Actually this is our interpretation of the concept of Gossamer superconductivity 
introduced by Bob 
Laughlin. As discussed in Ref. \cite{gossamer}, the case of Gossamer superconductivity in high $T_c$ cuprate 
superconductors, 
CeCoIn$_5$ and very likely $\kappa$-(ET)$_2$ salts is clear\cite{pinteric}.

In other words all the electronic ground states in crystalline solids belong to one of three possibilities: 

a. unconventional superconductivity

b. unconventional density wave and

c. coexistence of two of them. \\
These ground states have been expected from the renormalization group analysis of 2D 
and 
3D fermion systems. In these circumstances, the identification of the character and the symmetry of the order parameter 
is the first step. Now with ADMR, the giant Nernst effect and the nonlinear Hall coefficient, the exploration becomes 
much easier.

\section{Acknowledgements}

We have benefitted with collaborations and discussions with Carmen Almasan, Philipp Gegenwart, Tao Hu, Bojana 
Korin-Hamzi\'c, Peter Thalmeier and Silvia Tomi\'c.
B. D\'ora acknowledges the hospitality and support of the Max-Planck Institute for Chemical Physics of Solids, Dresden, 
where part of this work was done.
B. D\'ora was supported by the Magyary Zolt\'an postdoctoral
program of Magyary Zolt\'an Foundation for Higher Education (MZFK).
This work was supported by the Hungarian
Scientific Research Fund under grant numbers OTKA T037976, TS049881 and T046269.

 \bibliographystyle{apsrev}
 \bibliography{eth}
%


\printindex
\end{document}